\documentclass[prl,aps,preprintnumbers,twocolumn]{revtex4}
\usepackage{epsfig}
\usepackage{amsmath}

\begin{document}

\preprint{BNL-NT-06/11} \preprint{RBRC-588}

\title{A Unified Picture for Single Transverse-Spin Asymmetries in Hard Processes}

\author{Xiangdong Ji$^{1,2}$, Jian-Wei Qiu$^3$, Werner Vogelsang$^{4,5}$, Feng Yuan$^5$}
\affiliation{${}^1$Physics Department, University of Maryland,
College Park, MD 20742\\
${}^2$Physics Department, Peking University, Beijing 100871, P.R.
China\\
${}^3$Department of Physics and Astronomy, Iowa State University,
Ames, IA 50011\\
${}^4$Physics Department, Brookhaven National Laboratory, Upton,
NY 11973\\
${}^5$RIKEN BNL Research Center, Building 510A, Brookhaven
National Laboratory, Upton, NY 11973}
\date{\today}

\begin{abstract}
Using Drell-Yan pair production as an example, we explore the
relation between two well-known mechanisms for single
transverse-spin asymmetries in hard processes: twist-three
quark-gluon correlations when the pair's transverse momentum is
large, $q_\perp \gg \Lambda_{\rm QCD}$, and time-reversal-odd and
transverse-momentum-dependent parton distributions when $q_\perp$
is much less than the pair's mass. We find that although the two
mechanisms have their own domain of validity, they describe the
same physics in the kinematic region where they overlap. This
unifies the two mechanisms and imposes an important constraint on
phenomenological studies of single spin asymmetries.

\end{abstract}

\maketitle

\newcommand{\be}{\begin{equation}}
\newcommand{\ee}{\end{equation}}
\newcommand{\ben}{\[}
\newcommand{\een}{\]}
\newcommand{\beqn}{\begin{eqnarray}}
\newcommand{\eeqn}{\end{eqnarray}}
\newcommand{\Tr}{{\rm Tr} }

{\bf 1.} Single-transverse spin asymmetries (SSAs) have a long
history, starting from the 1970s and 1980s when large SSAs were first
observed in hadronic reactions at forward rapidities. The size of the
asymmetries came as a surprise and has posed a challenge for researchers
in Quantum Chromodynamics (QCD). SSAs
have again attracted much interest in recent years, both
experimentally and theoretically \cite{review}. Two mechanisms have been
proposed in QCD for explaining the large SSAs: effects due to
time-reversal-odd and transverse-momentum dependent parton
distributions (Sivers functions) \cite{Siv90}, or to
spin-dependent twist-three quark-gluon correlations
(Efremov-Teryaev-Qiu-Sterman (ETQS) mechanism) \cite{Efremov,qiu}.
Much progress has been made in understanding theoretical issues
involved in transverse-momentum-dependent (TMD) parton
distributions. Their gauge properties have been clarified
\cite{ColSop81,ColSopSte85,BroHwaSch02,Col02,BelJiYua02,BoeMulPij03},
and the relevant QCD factorization formulas have been established
\cite{ColSop81,ColSopSte85,JiMaYu04,ColMet04}. Phenomenological
studies have been made to explain the available data, using both
mechanisms\cite{qiu,siverscompare}.
An interesting relation between the two types of
distribution functions was discussed in \cite{BoeMulPij03}. However, a clear
connection between the two seemingly-different physical mechanisms,
particularly at the level of physical observables, has so far not been
established (see \cite {MaWa03}).

In this letter we explore the relation between the two mechanisms
in Drell-Yan pair production at transverse momentum $q_\perp$. If
the pair's mass is $Q$, the quark-gluon correlation mechanism
works when $ q_\perp, Q \gg \Lambda_{\rm QCD}$, the strong
interaction scale. On the other hand, the TMD QCD factorization
formalism applies when $q_\perp\ll Q$.  Therefore, one expects
there to be a common kinematic region, $\Lambda_{\rm QCD} \ll
q_\perp\ll Q$, where both mechanisms should apply and describe the
same physics. We perform calculations in this intermediate region
in both formalisms and find that they indeed yield the same
result. This conclusion imposes a rigorous constraint on
phenomenological studies of the single-spin asymmetry data.

Consider scattering of a transversely polarized proton of spin
$S_\perp$ and momentum $P$ on an unpolarized hadron (another
proton, for example) of momentum $P'$, producing a virtual photon
that subsequently decays into a pair of leptons. The total
center-of-mass energy is denoted by $s=(P+P')^2$. At the lowest
order in QCD perturbation theory, there are two ways to produce
such a pair: either a quark-antiquark pair annihilates into a
virtual photon after emitting a gluon of transverse momentum
$q_\perp$, or a quark scatters with a gluon producing a quark and
a virtual photon. If the lepton pair has a large positive
rapidity, $y$ (in the forward direction of the polarized proton),
the dominant contribution comes from scattering involving a quark
from the polarized proton. Our interest is to calculate the {\it
spin-dependent} differential cross section
$d\Delta\sigma(S_\perp)/dQ^2dyd^2q_\perp$ at large $y$ with
$\Delta\sigma(S_\perp)=[\sigma(S_\perp)-\sigma(-S_\perp)]/2$.

{\bf 2.} When $q_\perp$, $Q \gg \Lambda_{\rm QCD}$, the
spin-dependent cross section can be calculated in terms of a
twist-three quark-gluon correlation. The physics of this
correlation can be seen as follows: when a transversely-polarized
proton is traveling at nearly the speed of light, its internal color
electric and magnetic fields have preferred orientations in the
transverse plane. By parity invariance, the color electric
field must be orthogonal to the spin of the proton. If averaged
over the proton wave function, the field vanishes because the
proton is color-neutral (also because of time-reversal symmetry).
However, if one multiplies the color electric field with the quark color
current, the average may be non-zero. This average defines a
quark-gluon correlation function that characterizes a property
of a polarized proton.

High-energy scattering probes the so-called light-cone correlations.
For these, quark and/or gluon fields are separated along the light-cone
direction $\xi^-$ (if $\xi^\mu$ denotes a space-time coordinate,
the light-cone variables
are defined as $\xi^\pm = (\xi^0\pm\xi^3)/\sqrt{2}, \xi_\perp =
(\xi^1, \xi^2)$). A typical diagram for scattering of an antiquark in the
polarized gluon background is shown in Fig.~1. The blob in the lower
part of the diagram represents the spin-dependent quark-gluon
correlation described above. One finds the following expression for the
correlation \cite{qiu}:
\begin{eqnarray}
&&T_F(x_1,x_2) =
\int\frac{d\zeta^-d\eta^-}{4\pi}e^{i(k_{q1}^+\eta^-+k_g^+\zeta^-)}
\, \epsilon_\perp^{\beta\alpha}S_{\perp\beta}
\label{eq1} \\
&&\left\langle PS|\overline\psi(0){\cal L}(0,\zeta^-)\gamma^+
gF_{\alpha}^{\ +}(\zeta^-){\cal L}(\zeta^-,\eta^-)
\psi(\eta^-)|PS\right\rangle \nonumber \ ,
\end{eqnarray}
where the sums over color and spin indices are implicit,
$|PS\rangle$ denotes the proton state, $\psi$ the quark field, and
$F_{\alpha}^{\ +}$ the gluon field tensor. In Eq.~(\ref{eq1}),
$x_1=k_{q1}^+/P^+$ and $x_2=k_{q2}^+/P^+$ are the fractions of the
polarized proton's light-cone momentum carried by the quark in
Fig.~1, while $x_g=k_g^+/P^+ = x_2-x_1$ is the fractional momentum
carried by the gluon; ${\cal L}$ is the light-cone gauge link,
${\cal L}(\zeta_2,\zeta_1) = \exp\left(-ig\int^{\zeta_1}_{\zeta_2}
d\xi^- A^+(\xi^-)\right)$ that makes the correlation operator
gauge-invariant; and $\epsilon^{\alpha\beta}_\perp$ is the
2-dimensional Levi-Civita tensor with $\epsilon^{12}_\perp=1$.

\begin{figure}[t]
\begin{center}
\includegraphics[height=3.0cm]{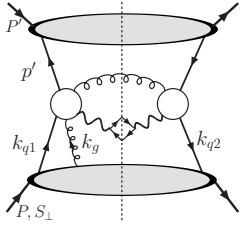}
\end{center}
\vskip -0.7cm \caption{\it A generic Feynman diagram contributing
to the single transverse-spin asymmetry for the Drell-Yan process
through the quark-antiquark scattering channel, containing both soft-
and hard-pole contributions. }
\end{figure}

When the antiquark of the unpolarized proton scatters off a quark
of the polarized proton in the presence of a polarized color
electromagnetic field (the gluon) in Fig.~1, the lepton-pair yield
has a dependence on the transverse orientation of the spin. The
strong interaction phase necessary for SSAs arises from the
interference between an imaginary part of the partonic scattering
amplitude with the extra gluon and a real part of the scattering
amplitude without a gluon in Fig.~1.  The imaginary part is from
the pole of the parton propagator associated with the integration
of the gluon momentum $x_g$. Depending on which propagator's pole
contributes, $\Delta\sigma(S_\perp)$ gets contributions from
$x_g=0$ (soft-pole) \cite{qiu} and $x_g\neq 0$ (hard-pole)
\cite{luo,xxx}.

{\bf 3.} By considering all partonic diagrams in Fig.~1 and their
complex conjugates, one finds for the spin-dependent differential
cross section:
\begin{eqnarray}
\frac{d^3\Delta\sigma^{q\bar q\to\gamma^*g}(S_\perp)}
{dQ^2dyd^2q_\perp}\!\!&\!\!=\!\!&\!\!\int\frac{dx}{x}\frac{dx'}{x'}
\bar q(x')\delta(\hat s+\hat t+\hat
u-Q^2)\nonumber\\
&&\times \,C\cdot \big[H_q^s+H_q^h\big](x,\hat{s},\hat{t},\hat{u}) \ ,
\end{eqnarray}
where the sum over all quark flavors, weighted with their electric
charge squared, is implicit. We have defined the factor $C=
\sigma_0\epsilon^{\alpha\beta}{S}_{\perp\alpha}
q_{\perp\beta}\alpha_s/2\pi^2$ with $\sigma_0 = 4\pi\alpha_{\rm
em}^2/3N_c sQ^2$, $N_c=3$. $x'$ is the momentum fraction carried by
the parton from the unpolarized proton, and $x$ is the total parton
momentum fraction from the polarized proton while $x_g$
is fixed using the pole (on-shell) condition. The partonic
Mandelstam variables are defined as $\hat s=(xP+x'P')^2$, $\hat
t=(xP-q)^2$, $\hat u=(x'P'-q)^2$. $H_q^s$ and
$H_q^h$ are the soft- and hard-pole contributions,
respectively.

The soft-pole contributions can be calculated in the same way as for
direct-photon and inclusive-hadron production \cite{qiu}.
A total of eight diagrams \cite{xxx} sum to
\begin{eqnarray}
H_q^s &=&\frac{1}{2N_c}\left[x\frac{\partial}{\partial
x}T_F(x,x)\frac{D^s_{q\bar q}}{-\hat u}+T_F(x,x) \frac{N^s_{q\bar
q}}{-\hat u}\right]\ , \label{hq}
\end{eqnarray}
where the hard coefficients $D^s_{q\bar q}$ and $N^s_{q\bar q}$ are
\begin{eqnarray} D^s_{q\bar q}&=&
\frac{\hat u}{\hat t}+\frac{\hat t}{\hat u}+\frac{2Q^2\hat s}{\hat
u\hat
t}\nonumber\\
N^s_{q\bar q}&=&\frac{1}{\hat t^2\hat u}\left[ Q^2(\hat u^2-\hat
t^2)+2Q^2\hat s(Q^2-2\hat
t)\nonumber\right.\\
&&\left. -(\hat u^2+\hat t^2)\hat t\right]\nonumber\ ,
\end{eqnarray}
with $1/2N_c$ the color factor. In the real photon limit,
$Q^2=0$, we obtain the direct-photon cross section.

Kinematics allows only a $t$-channel propagator to
contribute to a hard pole, and there
are a total of 12 such diagrams \cite{xxx}.
The sum of their contributions is
\begin{eqnarray}
H_q^h=T_F(x,x-x_g) \frac{N^h_{q\bar q}}{-\hat
u}\left[\frac{1}{2N_c}+C_F\frac{\hat s}{\hat s+\hat u}\right] \ ,
\end{eqnarray}
where
$$N^h_{q\bar q}=\frac{(Q^2-\hat t)^3+Q^2\hat
s^2}{-\hat t^2\hat u}\ . $$ As mentioned above, for the hard-pole
contributions $x_g=-x\hat t/(Q^2-\hat t)$ differs from zero. We
find that the hard-pole contribution has no ``derivative-term'',
$\propto \partial T_F(x,x)/\partial x$, due to a cancellation
among the diagrams.

{\bf 4.} When $q_\perp\ll Q$, single spin asymmetries can be
generated from a spin-dependent TMD quark distribution introduced
by Sivers \cite{Siv90}. The physics of the Sivers function can be
understood as follows. Consider a transversely-polarized proton
traveling with large momentum $P$. The distribution of quarks with
longitudinal and transverse momentum $xP$ and $\vec{k}_\perp$ can
have a dependence on the orientation of $\vec{k}_\perp$ relative
to the polarization vector $\vec{S}_\perp$. The TMD quark
distributions are best defined through the following matrix:
\begin{eqnarray}
    {\cal M}^{\alpha\beta} &=&   p^+\int
        \frac{d\xi^-}{2\pi}e^{-ix\xi^-P^+}\int
        \frac{d^2b_\perp}{(2\pi)^2} e^{i\vec{b}_\perp\cdot
        \vec{k}_\perp} \nonumber \\ && \times \left\langle
P\left|\overline{\mit \Psi}_v^\beta(\xi^-,0,\vec{b}_\perp) \gamma^+
        {\mit \Psi}_v^\alpha(0)\right|P\right\rangle\ .
\end{eqnarray}
Here the vector $p=(p^+,0^-,0_\perp)$ is along the momentum
direction of the proton, ${\mit \Psi}_v(\xi) = {\cal
L}_{v}(-\infty;\xi)\psi(\xi)$, and the gauge link is $ {\cal
L}_{v}(-\infty;\xi) = \exp\left(-ig\int^{-\infty}_0 d\lambda
v\cdot A(\lambda v +\xi)\right)$. In a system where $P^+\gg P^-$,
the vector $v$ with $v^-\gg v^+, v_\perp=0$ is given a non-zero
$v^+$ component in order to regulate the light-cone singularities.
An expansion of the matrix in Dirac indices yields
\begin{equation}
{\cal M} = \frac{1}{2}\left[q(x,k_\perp)\not\! p + q_T(x,k_\perp)
\epsilon_{\mu\nu\alpha\beta}\gamma^\mu p^\nu k^\alpha S^\beta +
...\right]
\end{equation}
where $q(x,k_\perp)$ is the spin-independent distribution and
$q_T(x, k_\perp)$ is the Sivers function. The dependence on $v$ is
of the form $\zeta^2=(2v\cdot P)^2/v^2$ and is implicit.

\begin{figure}[t]
\begin{center}
\includegraphics[height=3.0cm]{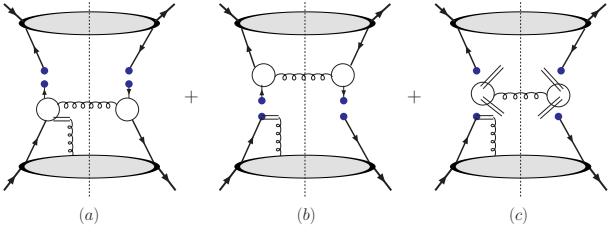}
\end{center}
\vskip -0.7cm \caption{\it Decomposition of the generic
diagram in Fig.~1 into different regions in the
transverse-momentum-dependent factorization approach. The large
transverse momentum of the Drell-Yan pair may come from (a) the
Sivers function, (b) the antiquark TMD distribution, and (c) the
soft factor. }
\end{figure}

When $q_\perp\ll Q$, the Drell-Yan cross section in leading order
$q_\perp/Q$ can be factorized in terms of TMD quark distributions.
A generic Feynman diagram can be decomposed into various
contributions, each of which is represented by one of the reduced
diagrams shown in Fig.~2 \cite{ColSopSte85}. This factorization
property can be seen from the diagram in Fig. 1 by considering the
real gluon emitted either along the direction of $P$ or $P'$. The
single-spin-dependent part of the Drell-Yan cross section has the
following factorization \cite{JiMaYu04},
\begin{eqnarray}
\frac{d^3\Delta\sigma(S_\perp)}{dQ^2dyd^2q_\perp}\!\!&\!=\!&\!\!
\sigma_0\frac{\epsilon^{\alpha\beta}{
S}_{\perp\alpha} q_{\perp\beta}}{M_P} \int
d^2\vec{k}_{1\perp}d^2\vec{k}_{2\perp}d^2\vec{\lambda}_\perp  H
\label{tmdfact} \\
&&\!\!\!\! \times \frac{\vec{k}_{1\perp}\cdot
\vec{q}_\perp}{\vec{q}_\perp^2}~\delta^{(2)}(\vec{k}_{1\perp}+
\vec{k}_{2\perp}+\vec{\lambda}_\perp-\vec{q}_\perp)
\nonumber\\
&&\!\!\!\! \times q_T(z_1,k_{1\perp},\zeta_1)~\bar
q(z_2,k_{2\perp},\zeta_2)~S(\lambda_\perp)^{-1} \ ,\nonumber
\end{eqnarray}
where $z_1=Q/\sqrt{s} e^y$ and $ z_2=Q/\sqrt{s} e^{-y}$ are the
momentum fractions of the colliding hadrons associated with the
observed Drell-Yan pair. $H$ is a hard factor and is entirely
perturbative. The soft-factor $S$ is a vacuum matrix element of
Wilson lines and captures the effect of soft gluon radiation.
Since the soft-gluon contribution in the TMD distributions has not
been subtracted, the soft-factor comes with inverse power. The
soft-gluon rapidity cut-off $\rho$ is defined as
$\rho=\sqrt{(2v_1\cdot v_2)^2/v_1^2v_2^2}$, where $v_1$ and $v_2$
are the directions of the gauge links in the Sivers and the
unpolarized antiquark distributions. In a special coordinate
frame, $z_1^2\zeta_1^2=z_2^2\zeta_2^2=\rho Q^2$ \cite{JiMaYu04}.

{\bf 5.} From what we have discussed so far, the two mechanisms
for the single-spin asymmetries appear to be quite different. In
the case of the twist-three correlation, the nonperturbative
function itself is time-reversal even. On the other hand, in the
TMD factorization, the Sivers function is time-reversal odd and
inherently contains the final-state interaction through a Wilson
line. There is, however, a common kinematic region, $\Lambda_{\rm
QCD} \ll q_\perp \ll Q$, where both mechanisms should work. In
this region, $q_\perp$ is large enough so that the asymmetry is a
twist-three effect. But at the same time, $q_\perp \ll Q$, so the
TMD factorization formalism also applies in this region.

In order to see the consistency of the two approaches, we need to
calculate the explicit $q_\perp$-dependence in the TMD
factorization. To do that, we let one of the transverse momenta
$\vec{k}_{i\perp}$ and $\vec{\lambda}_\perp$ be of the order of
$\vec{q}_\perp$ and the others small.  When $\vec{\lambda}_\perp$
is large, for example, we neglect $\vec{k}_{i\perp}$ in the delta
function, and the integrations over these momenta yield either the
ordinary antiquark distribution or a moment of the Sivers
function. The latter is related to the twist-three correlation
\cite{BoeMulPij03}:
\begin{eqnarray}
  \int d^2\vec{k}_\perp \bar q(x,k_\perp) = \bar q(x) \ ,
 \int d^2\vec{k}_\perp \frac{\vec{k}_\perp^2}{M_P} q_T(k_\perp, x) = T_F(x,x)\ . \nonumber
\end{eqnarray}
When one of the $\vec{k}_{i\perp}$ is taken to be of order
$\vec{q}_\perp$ and $\vec{\lambda}_\perp$ is neglected in the
delta function, one also needs $\int d^2\vec{\lambda}_\perp
S(\lambda_\perp) = 1$.

\begin{figure}[t]
\begin{center}
\includegraphics[height=3.0cm]{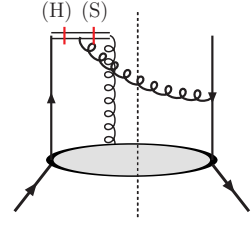}
\end{center}
\vskip -0.7cm \caption{\it A typical Feynman diagram contributing to
the Sivers functions at a large transverse momentum, $k_\perp \gg
\Lambda_{\rm QCD}$, arising from the twist-three quark-gluon
correlation. The bars indicate the soft and hard poles.}
\end{figure}

What remains is to calculate the large-$k_\perp$ behavior of the
TMD distributions and the soft factor.  The latter has been
determined in Ref. \cite{JiMaYu04} and reads
\begin{equation}
S(\lambda_\perp)=\delta^{(2)}(\vec{\lambda}_\perp) +
\frac{\alpha_s}{2\pi^2}\frac{1}{\vec{\lambda}_\perp^2}C_F\left(\ln
\rho^2-2\right) \ ,
\end{equation}
which is infra-red finite. To the same order, the unpolarized
antiquark TMD distribution is given by
\begin{eqnarray}
&& \bar q(z_2,k_\perp) = \bar q(z_2)\delta^{(2)}(\vec{k}_\perp)+
\frac{\alpha_s}{2\pi^2}\frac{1}{\vec{k}_\perp^2}C_F \\ && \times
\int\frac{dx'}{x'} \bar q(x')
\left[\frac{1+\xi_2^2}{(1-\xi_2)_+}+\delta(\xi_2-1)\left(\ln\frac{z_2^2\zeta_2^2}
{\vec{k}_\perp^2}-1\right)\right] \nonumber \ ,
\end{eqnarray}
where $\xi_2=z_2/x'$ and an additional contribution involving the gluon
distribution has been neglected \cite{xxx}.

As discussed above, the Sivers function at large $k_\perp$ can be
calculated in terms of the twist-three quark-gluon correlation. There are
also soft and hard pole contributions in this case. We show one example
of the relevant Feynman diagrams in Fig.~3.
There are a total of eight diagrams
for the soft-pole contribution, and twelve for the hard-pole one. Adding all
contributions, we find:
\begin{eqnarray}
q_T^{(1)}(z_1,k_\perp)&=&\frac{\alpha_s}{4\pi^2}\frac{2M_P}
{(\vec{k}_\perp^2)^2}\int\frac{dx}{x}
\left\{A+C_FT_F(x,x)\right.\nonumber\\
&&\times
\left.\delta(\xi_1-1)\left(\ln\frac{z_1^2\zeta_1^2}{\vec{k}_\perp^2}-
1\right)\right\} \ ,
\end{eqnarray}
where the $1/(k_\perp^2)^2$ behavior follows from a dimensional
analysis, and where
\begin{eqnarray} A\!\!&\!=\!&\!\! \frac{1}{2N_c}\left[
x\frac{\partial}{\partial
x}T_F(x,x)(1+\xi_1^2)+T_F(x,x-\tilde{x}_g)\frac{1+\xi_1}{(1-\xi_1)_+}\right.
\nonumber \\
&&\left.
+T_F(x,x)\frac{(1-\xi_1)^2(2\xi_1+1)-2}{(1-\xi_1)_+}\right]\nonumber\\
&&+C_FT_F(x,x-\tilde{x}_g)\frac{1+\xi_1}{(1-\xi_1)_+}
\end{eqnarray}
with $\tilde{x}_g=(1-\xi_1)x$ and $\xi_1=z_1/x$.

Plugging the above results into the TMD factorization formula~(\ref{tmdfact}),
we obtain the spin-dependent cross section  for the $q+\bar
q$ channel at large $q_\perp$:
\begin{eqnarray}
\frac{d^3\Delta\sigma^{q\bar q\to \gamma^*g}(S_\perp)
}{dQ^2dyd^2q_\perp}&=&\frac{C}{(q_\perp^2)^2}\int
\frac{dx}{x}\frac{dx'}{x'} \bar q(x') \label{tmd}\\
&&\times \left\{\delta(\xi_2-1)A+\delta(\xi_1-1)B\right\} \nonumber\
,
\end{eqnarray}
where $A$ and $C$ have been given above and
\begin{eqnarray}
B\!\!&\!=\!&\!\! C_FT_F(x,x)\left[\frac{1+\xi_2^2}{(1-\xi_2)_+}
+2\delta(\xi_2-1)\ln\frac{Q^2}{\vec{q}_\perp^{\,2}}\right] \, \nonumber
\end{eqnarray}
where $\xi_2=z_2/x'$.
To compare this to the twist-three correlation calculation, we
expand Eq.~(2) for small $q_\perp/Q$. At leading order in $q_\perp/Q$,
the partonic Mandelstam variables become $\hat
s={\vec{q}_\perp^{\,2}}/{(1-\xi_1)(1-\xi_2)}$, $ \hat
t=-{\vec{q}_\perp^{\,2}}/{(1-\xi_2)}$, and $ \hat
u=-{\vec{q}_\perp^{\,2}}/{(1-\xi_1)}$. Using the above, we find that the
spin-dependent cross section to leading order in $q_\perp/Q$ is exactly
the same
as the one shown in (\ref{tmd}). The same conclusion applies for the
contribution to the single-spin asymmetry generated by gluon-quark scattering.

To summarize, we have studied the single transverse-spin asymmetry
in Drell-Yan pair production at both large and small transverse
momenta $q_\perp$ of the lepton pair. At large $q_\perp$, one has
a result in terms of a twist-three quark-gluon correlation
function. At small $q_\perp$, a factorization formula in terms of
TMD parton distributions is valid. We have demonstrated that in
the intermediate region, both approaches give the same answer.
This establishes in an explicit calculation for a physical process
the connection between these two mechanisms for generating
single-spin asymmetries. It is this connection that unifies the
physical pictures for the underlying dynamics of single
transverse-spin asymmetries and imposes an important constraint on
phenomenological studies of the existing and future data.
Extension to the gluon initiated subprocess and other physical
processes, such as the semi-inclusive deep inelastic scattering,
will be presented elsewhere, along with the details of the
calculations described above.

We thank George Sterman for helpful discussion. X.~J. is supported
by the U. S. Department of Energy via grant DE-FG02-93ER-40762 and
by a grant from the Chinese National Natural Science Foundation
(CNSF). J.Q. is supported in part by the U. S. Department of
Energy under grant No. DE-FG02-87ER-40371. W.~V. and F.~Y. are
grateful to RIKEN, Brookhaven National Laboratory and the U.S.
Department of Energy (contract number DE-AC02-98CH10886) for
providing the facilities essential for the completion of their
work.

\end{document}